\documentclass[lettersize,journal]{IEEEtran}
\usepackage{amsmath,amsfonts}
\usepackage{algorithm}
\usepackage{array}
\usepackage[caption=false,font=normalsize,labelfont=sf,textfont=sf]{subfig}
\usepackage{textcomp}
\usepackage{stfloats}
\usepackage{url}
\usepackage{verbatim}
\usepackage{graphicx}
\usepackage{cite}
\usepackage{amssymb} 
\usepackage{xcolor}
\usepackage{algorithmicx}
\usepackage{algpseudocode}
\usepackage{tabularx}  
\usepackage{booktabs}  

\usepackage{hyperref} 
\hypersetup{
	colorlinks=true,       
	linkcolor=blue,         
	anchorcolor=black,     
	citecolor=blue,        
	urlcolor=blue,         
}
\begin{document}
	
	\title{Adaptive Structured Sparse Bayesian Learning for Near-Field Non-Stationary Channel Estimation in XL-MIMO Systems}
	\author{Qingxia Feng, Pan Fang, Meng Hua, Chunguo Li, Yongming Huang, and Luxi Yang
		
		
		}

\markboth{}%
{Shell \MakeLowercase{\textit{et al.}}: A Sample Article Using IEEEtran.cls for IEEE Journals}


\maketitle

\begin{abstract}
	Extremely large-scale multiple-input multiple-output (XL-MIMO) is a key enabler for sixth-generation (6G) communications. However, near-field channel estimation is particularly challenging due to spherical-wave propagation and spatial non-stationarity. To tackle this challenge, we propose a structured sparse Bayesian learning framework with adaptive dictionary updating for near-field non-stationary channel estimation. Specifically, the proposed method iteratively updates the distance parameters within an adaptive dictionary, thereby enhancing the representation capability without increasing the dictionary size. Moreover, we develop a hierarchical prior model that jointly captures polar-domain sparsity and structured dependency, enabling efficient Bayesian inference. Simulation results demonstrate that the proposed approach outperforms existing polar-domain dictionary-based methods while achieving low dictionary overhead.	
\end{abstract}


\section{Introduction}
\IEEEPARstart{D}{riven} by the demand for higher spectral efficiency and massive connectivity, extremely large-scale multiple-input multiple-output (XL-MIMO) and high-frequency transmissions have emerged as key enabling technologies for sixth-generation (6G) networks \cite{9390169}. The enlarged array apertures and short wavelengths extend the Rayleigh distance to several hundred meters \cite{9903389,10379539}, such that users may operate in the near-field region, where spherical-wave propagation must be explicitly considered. Moreover, the large aperture introduces pronounced spatial non-stationarity, such that the power contribution of a given path may vary significantly over antenna elements, yielding an effective visibility region (VR) that covers only a limited subset of antennas \cite{9940939,10295381}. Consequently, near-field XL-MIMO channel estimation is particularly challenging, as it must jointly account for spherical-wave propagation and spatial non-stationarity.

To tackle near-field non-stationary channel estimation, \cite{9777939} proposed a VR-aided framework that first identified the user VR by exploiting the statistical characteristics of the received power across antenna elements and then estimated the channel based on the detected VR. However, such power statistics were sensitive to noise and may lead to performance degradation,  particularly in low signal-to-noise ratio (SNR) regimes \cite{11300920}. Another approach in \cite{10373799} employed a group time block code (GTBC) to transform the non-stationary estimation task into several spatially stationary subproblems, but the spatial correlation across subarrays was not considered. More recently, \cite{10509715} developed a two-stage VR detection and channel estimation (TS-VDCE) scheme based on approximate message passing, which effectively exploited the spatial-domain characteristics of the channel but required manual separation between the two stages. In \cite{10634218}, a two-level orthogonal matching pursuit (T-OMP) algorithm was proposed to exploit the two-level sparsity of non-stationary channels; nevertheless, OMP-based methods typically required prior knowledge of the channel sparsity. These limitations motivated Bayesian modeling, since sparse Bayesian learning (SBL) generally improved robustness to unknown sparsity. Following this direction, \cite{10153711} proposed an SBL framework based on a polar-domain dictionary constructed via dense sampling in the angle–distance domain.

A common characteristic of the above schemes is that they rely on a predefined dictionary for sparse representation, making the estimation performance highly dependent on the dictionary design. In particular, most existing studies adopt either a fixed discrete Fourier transform (DFT) dictionary \cite{11300920,10509715} or a polar-domain dictionary \cite{9777939,10373799,10634218,10153711}. The DFT dictionary does not encode distance information, which limits its sparse representation capability for near-field channel estimation \cite{Cui2022_Channel}. In contrast, the polar-domain dictionary depends on two-dimensional sampling in the angle-distance domain. This discretization makes the estimation accuracy dependent on the sampling resolution, and achieving high accuracy typically requires an excessively large dictionary, which results in high computational complexity.

Motivated by these limitations, we propose an adaptive structured sparse Bayesian learning algorithm with dictionary updating (ASSBL) for near-field non-stationary channel estimation, aiming to reduce the dictionary overhead of conventional polar-domain schemes. The proposed method first extracts coarse angular support and then iteratively refines distance parameters in the dictionary, enabling a flexible near-field representation without enlarging the dictionary size. To account for spatial non-stationarity, we introduce a spatial visibility vector and develop a hierarchical prior that jointly captures polar-domain sparsity and the structured dependency. Under this prior, the channel estimation problem is cast as structured sparse reconstruction, and an efficient Bayesian inference algorithm is derived. Numerical results demonstrate that ASSBL achieves superior estimation performance with reduced dictionary overhead.
\section{System Model and Problem Formulation}
\begin{figure}[!t]
	\centering
	\includegraphics[width=2.9in]{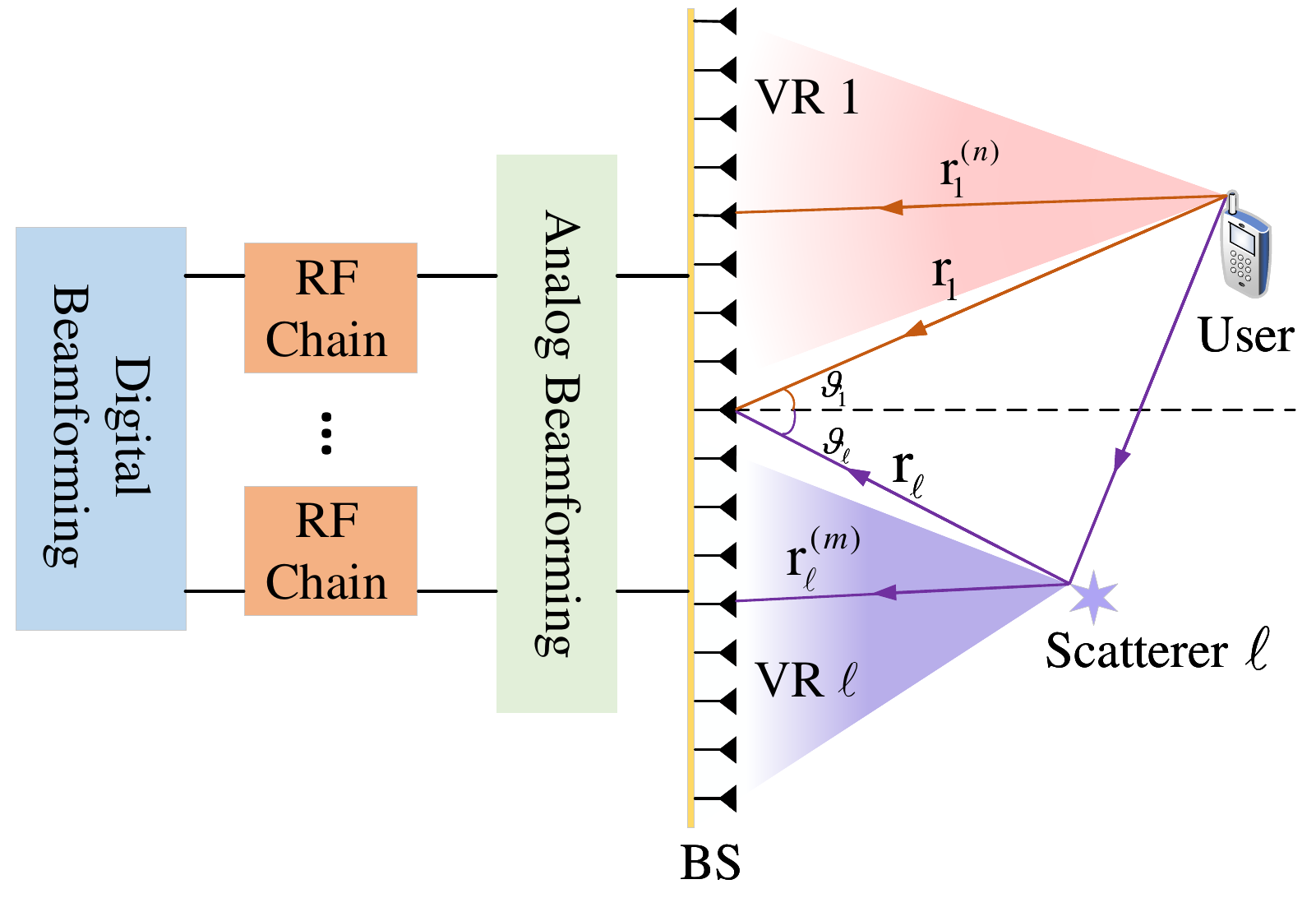}
	\caption{The near-field non-stationary channel model.}
	\label{fig_1}
\end{figure}
\subsection{System Model}
We consider an uplink time-division duplexing (TDD)-based XL-MIMO system, where the base station (BS) is equipped with a uniform linear array (ULA) of $N$ antenna elements with half-wavelength spacing $d=\lambda_c/2$, and the $n$-th antenna element is located at $(0,\delta_n d)$ with $\delta_n=\frac{2n-N-1}{2}$. The BS adopts a hybrid combining architecture with $N_{\mathrm{RF}}$ radio-frequency (RF) chains. 

During the uplink training phase, orthogonal pilot symbols are transmitted over $T_p$ time slots and combined at the BS using analog combiners subject to constant-modulus constraints. Without loss of generality, we assume unit-modulus pilot symbols, i.e., $s_t=1$ for $t=1,\ldots,T_p$. After stacking the received signals over all training slots, the aggregated observation can be expressed as
\vspace{-2mm}
\begin{equation}
	\mathbf{y}=\mathbf{W}^H\mathbf{h}+\mathbf{n},
	\label{eq:aggregated_signal}
\end{equation}
where $\mathbf{h}\in\mathbb{C}^{N\times1}$ denotes the uplink channel vector, $\mathbf{W}=[\mathbf{W}_{1},\ldots,\mathbf{W}_{T_p}]\in\mathbb{C}^{N\times T_pN_{\mathrm{RF}}}$ is the aggregated analog combining matrix with $|[\mathbf{W}_t]_{j,k}|=1/\sqrt{N}$, and $\mathbf{n}\in\mathbb{C}^{T_pN_{\mathrm{RF}}\times1}$ represents the combined additive white Gaussian noise.

\subsection{Near-Field Spatial Non-stationary Channel Model}
Based on a geometric multipath propagation model \cite{Cui2022_Channel}, the near-field uplink channel is expressed as
\vspace{-2mm}
\begin{equation}
	\mathbf{h} = \sqrt{\frac{N}{L}} \sum_{\ell=1}^{L} \alpha_\ell   \mathbf{a}(\theta_\ell, r_\ell) \odot \mathbf{b}_\ell,
\end{equation}
where $L$ denotes the number of propagation paths, $\alpha_\ell$ is the complex gain of the $\ell$-th path, where $\theta_\ell \in [-1, 1]$ and $r_\ell$ denote the normalized spatial angle and distance, respectively. The vector $\mathbf{b}_\ell \in \{0, 1\}^{N \times 1}$ characterizes the spatial visibility of the $\ell$-th path. The near-field steering vector is given by
\vspace{-2mm}
\begin{equation}
	\mathbf{a}(\theta_\ell,r_\ell)=\frac{1}{\sqrt{N}}[e^{-j\frac{2\pi}{\lambda}(r_\ell^{(1)}-r_\ell)},\cdots,e^{-j\frac{2\pi}{\lambda}(r_\ell^{(N)}-r_\ell)}]^T,
\end{equation}
where the propagation distance to the $n$-th antenna is approximated by $r_\ell^{(n)} \approx r_\ell - \delta_n d \theta_\ell + \frac{\delta_n^2 d^2 (1 - \theta_\ell^2)}{2 r_\ell}$, capturing spherical-wave propagation.

In near-field propagation, the channel exhibits spatial non-stationarity, which can be characterized by the VR of each propagation path \cite{10295381}. To facilitate tractable analysis and algorithm design, the ULA is partitioned into $G$ sub-arrays, each consisting of $N_g = N/G$ consecutive antenna elements. Accordingly, the visibility vector can be expressed as $\mathbf{b}_{\ell}=\mathbf{J}\bar{\mathbf{b}}_{\ell}$, where $\bar{\mathbf{b}}_{\ell}=[\bar{b}_{{\ell},1},\ldots,\bar{b}_{{\ell},G}]^T\in\{0,1\}^{G \times 1}$. $\mathbf{J}=\mathbf{I}_G\otimes\mathbf{1}_{N_g}\in\mathbb{R}^{N\times G}$ is the sub-array expansion matrix. Here, $\bar{b}_{{\ell},g}=1$ indicates that the $g$-th sub-array lies within the VR of the $\ell$-th path, and $\bar{b}_{{\ell},g}=0$ otherwise.
 
\subsection{Structured Sparse Representation in the Polar-Domain}
To exploit the sparsity of near-field channels in the polar domain, the channel is represented using a polar-domain dictionary defined as
\vspace{-2mm}
\begin{equation}
 	\mathbf{A}=[\mathbf{a}(\theta_1,r_1),\ldots,\mathbf{a}(\theta_1,r_M),\ldots,\mathbf{a}(\theta_{U}, r_{1}),\ldots,\mathbf{a}(\theta_U,r_{M})],
\end{equation}
where $\theta_u=\frac{(2u-U-1)}{U}$, $u=1,\ldots,U$, denotes the $u$-th uniformly sampled angular grid point, and $M$ is the number of sampled points along the distance ring. For notational convenience, the polar-domain dictionary is compactly written as $\mathbf{A}=[\mathbf{a}_1,\ldots,\mathbf{a}_Q]\in\mathbb{C}^{N\times Q}$, where $Q=UM$ denotes the total number of sampled near-field atoms.
 
Accordingly, the near-field non-stationary channel can be sparsely represented as
\vspace{-2mm}
\begin{equation}
	\begin{aligned}
		\mathbf{y}
		&=\mathbf{W}^H\sum_{q=1}^Q x_q\big(\mathbf{a}_q\odot\mathbf{b}_{q}\big)+\mathbf{n}
		=\mathbf{W}^H{\boldsymbol{\Phi}}{\bar{\mathbf{z}}}+\mathbf{n},
	\end{aligned}
\end{equation}
where $\mathbf{b}_q=\mathbf{J}\bar{\mathbf{b}}_q$ denotes the visibility pattern of the $q$-th atom over the array. Moreover,
$\bar{\mathbf{z}}=\operatorname{vec}\!\big(\mathbf{B}\operatorname{diag}(\mathbf{x})\big)
=[\mathbf{z}_1^T,\ldots,\mathbf{z}_Q^T]^T\in\mathbb{C}^{GQ\times1}$,
where $\mathbf{z}_q=x_q\bar{\mathbf{b}}_{q}\in\mathbb{C}^{G\times1}$ denotes the structured sparse coefficient associated with the $q$-th polar-domain atom, and
$\mathbf{B}=[\bar{\mathbf{b}}_1,\ldots,\bar{\mathbf{b}}_Q]\in\{0,1\}^{G\times Q}$ is the visibility matrix. The corresponding structured polar-domain dictionary is defined as $\boldsymbol{\Phi}=[\boldsymbol{\Phi}_1,\ldots,\boldsymbol{\Phi}_Q]\in\mathbb{C}^{N\times GQ}$,
where $\boldsymbol{\Phi}_q=\mathrm{diag}(\mathbf{a}_q)\mathbf{J}\in\mathbb{C}^{N\times G}$.
The coefficient vector $\mathbf{x}=[x_1,\ldots,x_Q]^T\in\mathbb{C}^{Q\times1}$ is sparse, where each nonzero entry represents the complex gain of a dominant polar-domain atom.

In this formulation, the vector $\bar{\mathbf{z}}$ exhibits a hierarchical sparsity structure inherent to near-field XL-MIMO propagation. Specifically, only a small number of blocks $\{\mathbf{z}_q\}$ are active, reflecting the limited number of dominant scatterers. Within each active block, the coefficients capture the visibility pattern of the corresponding atom across the array. Although such structured sparsity enables efficient reconstruction, a direct polar-domain discretization typically leads to an overcomplete dictionary, thereby incurring high computational complexity. This motivates the adoption of an adaptive dictionary within a structured sparse Bayesian learning framework to preserve the hierarchical sparsity structure while reducing the effective dictionary size.

\section{Proposed ASSBL Channel Estimation with Adaptive Dictionary}

\subsection{Adaptive Dictionary Construction}
Conventional far-field channel estimation commonly relies on an angle-domain DFT dictionary, which ignores range information and becomes inadequate in near-field scenarios. The polar-domain dictionary, incorporating both angle and distance information, improves modeling accuracy but grows rapidly with sampling resolution, leading to increased complexity. To reduce the dictionary overhead, we construct a distance-parameterized adaptive dictionary that preserves the angular structure and iteratively refines the distance parameters, avoiding exhaustive discretization of the angle–distance domain. Specifically, the near-field channel is modeled as
\vspace{-2mm}
\begin{equation}
	\mathbf{h}=\mathbf{D}(\mathbf{r}){\mathbf{z}},
\end{equation}
where $\mathbf{D}(\mathbf{r})=[\mathbf{D}_1,\ldots,\mathbf{D}_U]\in\mathbb{C}^{N\times GU}$ denotes the adaptive dictionary, and $\mathbf{z}=[\mathbf{z}_1^T,\ldots,\mathbf{z}_U^T]^T\in\mathbb{C}^{GU\times1}$ is the structured coefficient vector. The $u$-th sub-dictionary is defined as $\mathbf{D}_u=\mathrm{diag}(\mathbf{a}_u)\mathbf{J}\in\mathbb{C}^{N\times G}$, where $\mathbf{a}_u=\mathbf{a}(\theta_u,r_u)$ is the near-field steering vector associated with the sampled angle $\theta_u$ and distance $r_u$. The angular and distance-parameter are denoted by $\boldsymbol{\theta}=\{\theta_1,\ldots,\theta_U\}$ and $\mathbf{r}=\{r_1,\ldots,r_U\}$, respectively.

The adaptive dictionary $\mathbf{D}(\mathbf{r})$ requires refining the distance parameters $\mathbf{r}$ by leveraging measurement information across angular components. Specifically, coarse angular information is first used to identify dominant directions, after which the associated distances are updated through dictionary refinement. Within the proposed ASSBL framework, the structured coefficients and distance parameters are jointly inferred via Bayesian learning, leading to a compact near-field channel estimator.

\subsection{Structured Hierarchical Sparse Bayesian Framework}
To further exploit the structure of $\mathbf{z}=[\mathbf{z}_1^T,\ldots,\mathbf{z}_U^T]^T$, we adopt a hierarchical sparse prior given by
\vspace{-2mm}
\begin{equation}
	p(\mathbf{z}|\boldsymbol{\gamma},\boldsymbol{\Delta})
	=\prod_{u=1}^U p(\mathbf{z}_u|\gamma_u,\boldsymbol{\Delta}_u),
\end{equation}
where $\boldsymbol{\gamma}=[\gamma_1,\ldots,\gamma_U]^T$ denotes the block-level relevance hyperparameters, and $\boldsymbol{\Delta}=\{\boldsymbol{\Delta}_1,\ldots,\boldsymbol{\Delta}_U\}$ characterizes the intra-block covariance structure. As $\gamma_u$ approaches zero, the corresponding block $\mathbf{z}_u$ is effectively suppressed, resulting in block-level sparsity  \cite{6967808}. Conditioned on the hyperparameters, each block $\mathbf{z}_u$ is assumed to follow a complex Gaussian distribution
\vspace{-2mm}
\begin{equation}
	p(\mathbf{z}_u|\gamma_u,\boldsymbol{\Delta}_u)
	=\mathcal{CN}(\mathbf{z}_u|\mathbf{0},\gamma_u\boldsymbol{\Delta}_u).
\end{equation}

To promote structured sparsity within the Bayesian framework, the block-level hyperparameters $\{\gamma_u\}$ are assigned Gamma priors 
\vspace{-2mm}
\begin{equation}
	p(\gamma_u|\lambda)=\Gamma(\gamma_u;1,\lambda),
\end{equation}
where $\lambda$ denotes the rate parameter and $p(\lambda)=\Gamma(\lambda;a_\lambda,b_\lambda)$. The block-internal covariance matrix is modeled as $\boldsymbol{\Delta}_u=\mathrm{diag}(\delta_{u,1},\ldots,\delta_{u,G})$, with $\delta_{u,g}=(\alpha_{u,g}+\beta\alpha_{u,g-1}+\beta\alpha_{u,g+1})^{-1}$, where $\{\alpha_{u,g}\}$ are nonnegative hyperparameters controlling the local structure within each block, $\beta\geq 0$ regulates the coupling strength between adjacent coefficients, and the boundary conditions $\alpha_{u,0}=\alpha_{u,G+1}=0$ are adopted \cite{6967808}. The vector $\boldsymbol{\alpha}_u=[\alpha_{u,1},\ldots,\alpha_{u,G}]^{T}$ is assigned Gamma priors to promote sparsity in a hierarchical Bayesian manner, given by
\vspace{-2mm}
\begin{equation}
p(\alpha_{u,g}|\zeta_u)=\Gamma(\alpha_{u,g};1,\zeta_u),
\end{equation}
where $\{\zeta_u\}$ are higher-level hyperparameters, which are further assigned Gamma priors $p(\zeta_u)=\Gamma(\zeta_u;a_\zeta,b_\zeta)$.

In addition to the channel coefficients, the noise precision parameter $\sigma$ is treated as an unknown random variable and assigned a Gamma prior given by
\vspace{-2mm}
\begin{equation}
	p(\sigma)=\Gamma(\sigma;a_\sigma,b_\sigma).
\end{equation}

Given the sparse coefficient vector $\mathbf{z}$ and the noise precision parameter $\sigma$, the observation model is expressed as
\vspace{-2mm}
\begin{equation}
	p(\mathbf{y}|\mathbf{z},\sigma)
	=\mathcal{CN}(\mathbf{y}|\boldsymbol{\Psi}\mathbf{z},\sigma^{-1}\mathbf{I}_{T_pN_{\mathrm{RF}}}),
\end{equation}
where $\boldsymbol{\Psi}=\mathbf{W}^H\mathbf{D}(\mathbf{r})$ denotes the effective sensing matrix. The resulting joint distribution factorizes as
\vspace{-2mm}
\begin{equation}
	\begin{aligned}
		p(\mathbf{y},\mathbf{z},\boldsymbol{\gamma},\boldsymbol{\alpha},\boldsymbol{\zeta},\lambda,\sigma)=&p(\mathbf{y}|\mathbf{z},\sigma) p(\mathbf{z}|\boldsymbol{\gamma},\boldsymbol{\Delta}) p(\boldsymbol{\gamma}|\lambda)\\&\cdot
		p(\boldsymbol{\alpha}|\boldsymbol{\zeta}) p(\boldsymbol{\zeta}) p(\lambda) p(\sigma).
	\end{aligned}
\end{equation}

\subsection{Adaptive Inference of Unknown Parameters}
Under the hierarchical prior described above, estimating $\mathbf{z}$ becomes a structured sparse inference problem. We aim to jointly infer $\mathbf{z}$ and the hyperparameter set
$\boldsymbol{\Theta}=\{\boldsymbol{\gamma},\boldsymbol{\alpha},\boldsymbol{\zeta},\lambda,\sigma\}$. To this end, an EM-type iterative inference procedure is adopted, which alternates between posterior estimation and parameter updates \cite{4644060}. Given the current hyperparameter estimates, the posterior distribution of $\mathbf{z}$ in the E-step follows a complex Gaussian distribution expressed as
\vspace{-2mm}
\begin{equation}
	p(\mathbf z|\mathbf y;\boldsymbol\Theta)=\mathcal{CN}(\mathbf z|\boldsymbol\mu,\boldsymbol\Sigma),
\end{equation}
where the posterior mean and covariance are given by
\vspace{-2mm}
\begin{equation}
	\boldsymbol\mu=\sigma\boldsymbol\Sigma \boldsymbol\Psi^H\mathbf y,\quad
	\boldsymbol\Sigma=(\sigma\boldsymbol\Psi^H\boldsymbol\Psi+\boldsymbol\Omega^{-1})^{-1},\label{cova}
\end{equation}
with $\boldsymbol{\Omega}=\mathrm{blkdiag}\left(\gamma_1\boldsymbol{\Delta}_1,\ldots,\gamma_U\boldsymbol{\Delta}_U\right)$.

In the M-step, the hyperparameters and dictionary parameters are updated as 
\vspace{-1mm}
\begin{equation}
(\boldsymbol{\Theta}^{(i+1)},\mathbf{D}^{(i+1)})
	=\arg\max_{(\boldsymbol{\Theta},\mathbf{D})}\, \mathcal Q(\boldsymbol{\Theta},\mathbf{D}|\boldsymbol{\Theta}^{(i)},\mathbf{D}^{(i)}),
\end{equation}
where $\mathcal{Q}(\boldsymbol{\Theta},\mathbf{D})=\mathbb{E}_{p(\mathbf z|\mathbf y;\boldsymbol\Theta)}\left[\ln p(\mathbf{y},\mathbf{z},\boldsymbol{\Theta})\right]$ and $i$ denotes the iteration index.

\subsubsection{Updating $\{\gamma_u\}$}
For each block-level hyperparameter $\gamma_u$, the objective function is given by
\vspace{-1mm}
\begin{equation}
	\begin{aligned}
	\mathcal Q(\gamma_u)=-\frac{\sum_{g=1}^G\delta_{u,g}^{-1}(|\mu_{u,g}|^2+[\boldsymbol\Sigma_u]_{g})}{\gamma_u}-G\ln\gamma_u-\lambda\gamma_u,\label{Q_gamma}
	\end{aligned}
\end{equation}
where $[\boldsymbol\Sigma_u]_{g}$ represents the posterior variance of the $g$-th coefficient within the $u$-th block. Taking the derivative of \eqref{Q_gamma} with respect to $\gamma_u$ and setting it to zero yields
\vspace{-1mm}
\begin{equation}
	\gamma_u^{(i+1)}=		
	\frac{-G+\sqrt{G^2+4\lambda \sum_{g=1}^G\delta_{u,g}^{-1}(|\mu_{u,g}|^2+[\boldsymbol\Sigma_u]_{g})}}{2\lambda}.\label{rule_gamma}
\end{equation}

\subsubsection{Updating $\{ \boldsymbol{\alpha}_u \}$}
The intra-block hyperparameters $\{ \boldsymbol{\alpha}_u \}$ are updated in the next step. By retaining only the terms dependent on $\alpha_{u,g}$, the objective function becomes
\begin{equation}
	\begin{aligned}
	\mathcal{Q}(\alpha_{u,g})
	=&-\frac{1}{\gamma_{u}}(\beta (|\mu_{u,g-1}|^2+[\boldsymbol\Sigma_u]_{g-1})+(|\mu_{u,g}|^2+[\boldsymbol\Sigma_u]_{g})\\&+\beta (|\mu_{u,g+1}|^2+[\boldsymbol\Sigma_u]_{g+1}))\alpha_{u,g}+(\ln\delta_{u,g-1}^{-1}\\&+\ln\delta_{u,g}^{-1}+\ln\delta_{u,g+1}^{-1})
	-\zeta_{u}\alpha_{u,g}.\label{Q_alpha}
	\end{aligned}
\end{equation}
Taking the derivative of \eqref{Q_alpha} with respect to $\alpha_{u,g}$ and setting it to zero yields
\begin{equation}
	\alpha_{u,g}^{(i+1)}=\chi(1+2\beta)({\gamma_u}^{-1}\nu_{u,g}+\zeta_u)^{-1},\label{rule_alpha}
\end{equation}
where $\chi\in(0,1)$ and $\nu_{u,g}=|\mu_{u,g}|^2+[\boldsymbol\Sigma_u]_{g}+\beta (|\mu_{u,g-1}|^2+[\boldsymbol\Sigma_u]_{g-1})+\beta (|\mu_{u,g+1}|^2+[\boldsymbol\Sigma_u]_{g+1})$.

\subsubsection{Updating $\{ \zeta_u \}$}By retaining only the terms dependent on $\zeta_u$, the objective function reduces to
\vspace{-2mm}
\begin{equation}
	\begin{aligned}
	\mathcal{Q}(\zeta_u)&
	=-\zeta_u(\sum_{g=1}^G\alpha_{u,g}+b_\zeta)+(G+a_\zeta-1)\ln\zeta_u.\label{Q_zeta}
	\end{aligned}
\end{equation}
Taking the derivative of \eqref{Q_zeta} with respect to $\zeta_u$ and setting it to zero yields
\begin{equation}
	\zeta_u^{(i+1)}=		
	\frac{G+a_\zeta-1}{\sum_{g=1}^G\alpha_{u,g}+b_\zeta}.\label{rule_zeta}
\end{equation}

\subsubsection{Updating $\lambda$}
Next, by retaining only the terms dependent on $\lambda$, the objective function reduces to
\vspace{-2mm}
\begin{equation}
	\begin{aligned}
	\mathcal{Q}(\lambda)&=-\lambda(\sum_{u=1}^U\gamma_u+b_\lambda)+(U+a_\lambda-1)\ln\lambda.\label{Q_lambda}
	\end{aligned}
\end{equation}
Taking the derivative of \eqref{Q_lambda} with respect to $\lambda$ and setting it to zero yields
\vspace{-2mm}
\begin{equation}
	\lambda^{(i+1)}=		
	\frac{U+a_\lambda-1}{\sum_{u=1}^U\gamma_u+b_\lambda}.\label{rule_lambda}
\end{equation}

\subsubsection{Updating $\sigma$}
Then, by retaining only the terms dependent on $\sigma$, the objective function reduces to
\vspace{-2mm}
\begin{equation}
	\begin{aligned}
	\mathcal{Q}(\sigma)	=&(T_pN_{\mathrm{RF}}+a_\sigma-1)\ln\sigma\\&
	-\sigma(\|\mathbf y-\boldsymbol\Psi\boldsymbol\mu\|_2^2
	+\mathrm{tr}(\boldsymbol\Psi^H\boldsymbol\Psi\boldsymbol\Sigma)+b_\sigma).\label{Q_sigma}
	\end{aligned}
\end{equation}
Taking the derivative of \eqref{Q_sigma} with respect to $\sigma$ and setting it to zero yields
\vspace{-2mm}
\begin{equation}
	\sigma^{(i+1)}=
	\frac{T_pN_{\mathrm{RF}}+a_\sigma-1}
	{\|\mathbf y-\boldsymbol\Psi\boldsymbol\mu\|_2^2
		+\mathrm{tr}(\boldsymbol\Psi^H\boldsymbol\Psi\boldsymbol\Sigma)+b_\sigma}.\label{rule_sigma}
\end{equation}

\subsubsection{Updating $\mathbf{D(\mathbf{r})}$}
Next, the adaptive dictionary $\mathbf{D(\mathbf{r})}$ is refined by updating the associated distance parameters $\mathbf{r}$. Note that the posterior mean $\boldsymbol\mu$ provides an estimate of the sparse beamspace channel coefficients, while the covariance $\boldsymbol\Sigma$ captures the corresponding uncertainty under the current dictionary. Accordingly, the objective function becomes
\vspace{-2mm}
\begin{equation}
	\mathcal{Q}(\mathbf{D(\mathbf{r})})=\sigma\|\mathbf{y}-\boldsymbol{\Psi}(\mathbf{r})\boldsymbol{\mu}\|_2^2+\sigma\mathrm{tr}(\boldsymbol{\Psi}(\mathbf{r})\boldsymbol{\Sigma}\boldsymbol{\Psi}^H(\mathbf{r})).
\end{equation}

In general, a closed-form solution for $\mathbf{D(\mathbf{r})}$ is not available. Hence, an iterative refinement procedure is employed to update the dictionary. Specifically, the distance parameters are updated through an iterative rule with respect to $1/r_{u}$, as
\vspace{-2mm}
\begin{equation}
	\frac{1}{r_u^{(i+1)}}=\frac{1}{r_u^{(i)}}-\eta\nabla_{\frac{1}{r_u}}\mathcal{Q}(\mathbf{D}(r_u^{(i)})),\label{rule_r}
\end{equation}
where $\eta$ is the step size determined by the Armijo backtracking line search. The corresponding gradient is computed as
\vspace{-2mm}
\begin{equation}
	\frac{\partial \mathcal{Q}(\mathbf{D(\mathbf{r})})}{\partial \frac{1}{r_u}}=2\Re\{\mathrm{tr}([\sigma\left((\boldsymbol{\Psi\mu}-\mathbf{y})\boldsymbol{\mu}^{H}+\boldsymbol{\Psi\Sigma}\right)]_u^{H}\frac{\partial\boldsymbol{\Psi}}{\partial\frac{1}{r_u}})\},
\end{equation}
where $\frac{\partial\boldsymbol{\Psi}}{\partial\frac{1}{r_u}}=\mathbf{W}^{H}\mathrm{diag}(\frac{\partial\mathbf{a}_{u}}{\partial\frac{1}{r_u}})\mathbf{J}$ and $[\cdot]_u$ denotes the submatrix corresponding to the $u$-th block, i.e., the columns indexed by $(u-1)G+1:uG$ elements. 

The proposed algorithm is summarized in Algorithm~\ref{alg1}. Note that refining all distance parameters at every iteration is unnecessary, since atoms with negligible posterior power are unlikely to be activated. Therefore, only the $\tilde{U}$ most significant components are selected for distance refinement (e.g., $\tilde{U}=2L$).

\begin{algorithm}[t]
	\caption{The Proposed ASSBL Algorithm}
	\label{alg1}
	\begin{algorithmic}[1]
		
		\Statex \textbf{Input:} Received signal $\mathbf{y}$, combining matrix $\mathbf{W}$.
		\Statex \textbf{Repeat:}
		
		\State \parbox[t]{\dimexpr\linewidth-\algorithmicindent\relax}{%
			\hangindent=\algorithmicindent \hangafter=1
			Update the ${\boldsymbol\mu}^{(i+1)}$ and ${\boldsymbol\Sigma}^{(i+1)}$ according to \eqref{cova}.%
		}
		\State \parbox[t]{\dimexpr\linewidth-\algorithmicindent\relax}{%
			\hangindent=\algorithmicindent \hangafter=1
			Update the ${ \boldsymbol{\gamma}}^{(i+1)}$ according to \eqref{rule_gamma}.%
		}
		\State \parbox[t]{\dimexpr\linewidth-\algorithmicindent\relax}{%
			\hangindent=\algorithmicindent \hangafter=1
			Update the ${\boldsymbol{\alpha}_u}^{(i+1)}$ according to \eqref{rule_alpha}.%
		}
		\State \parbox[t]{\dimexpr\linewidth-\algorithmicindent\relax}{%
			\hangindent=\algorithmicindent \hangafter=1
			Update the ${\zeta_u}^{(i+1)}$ according to \eqref{rule_zeta}.%
		}
		\State \parbox[t]{\dimexpr\linewidth-\algorithmicindent\relax}{%
			\hangindent=\algorithmicindent \hangafter=1
			Update the ${\lambda}^{(i+1)}$ according to \eqref{rule_lambda}.%
		}
		\State \parbox[t]{\dimexpr\linewidth-\algorithmicindent\relax}{%
			\hangindent=\algorithmicindent \hangafter=1
			Update the ${\sigma}^{(i+1)}$ according to \eqref{rule_sigma}.%
		}
		\State \parbox[t]{\dimexpr\linewidth-\algorithmicindent\relax}{%
			\hangindent=\algorithmicindent \hangafter=1
			Update the first $\tilde{U}$ distances ${r_u}^{(i+1)}$ according to \eqref{rule_r}.%
		}
		\State \parbox[t]{\dimexpr\linewidth-\algorithmicindent\relax}{%
			\hangindent=\algorithmicindent \hangafter=1
			Update the adaptive dictionary ${\mathbf{D(\mathbf{r})}}^{(i+1)}$.%
		}
		
		\Statex \textbf{Until:} increase $i$ until $i \le I$ \textbf{or} $\left\|\boldsymbol\alpha^{(i+1)}-\boldsymbol\alpha^{(i)}\right\|_2\leq\epsilon$.
		
		\Statex \textbf{Output:} ${\mathbf{D(\mathbf{r})}}$ and ${\boldsymbol\mu}$.
		
	\end{algorithmic}
\end{algorithm}

\subsection{Computational Complexity}
The complexity of Algorithm~\ref{alg1} is dominated by computing the posterior mean $\boldsymbol\mu$ and covariance $\boldsymbol\Sigma$, which requires forming $\boldsymbol\Psi^{H}\boldsymbol\Psi$ and inverting a $K\times K$ matrix, leading to $\mathcal{O}(MK^{2}+K^{3})$, where $K=GU$ and $M=T_pN_{\mathrm{RF}}$. 
When $M\ll K$, by applying the Woodbury identity, the inversion is shifted to an $M\times M$ matrix and the complexity reduces to $\mathcal{O}(M^{2}K+M^{3})$. 
In addition, updating the adaptive dictionary over $\tilde{U}$ dominant components incurs $\mathcal{O}(\tilde{U}MG)$ complexity per iteration. 
Therefore, when $M\ll K$, the overall complexity over $I$ iterations is approximately 
$\mathcal{O}\!(I(M^{2}K+M^{3}+\tilde{U}MG))$.

\section{Numerical Results}
In this section, simulation results are provided to evaluate the performance of the proposed near-field channel estimation scheme. The normalized mean square error (NMSE) is adopted as the performance metric, defined as $\mathrm{NMSE}=\mathbb{E}\!\left[\frac{\|\mathbf{h}-\hat{\mathbf{h}}\|_2^2}{\|\mathbf{h}\|_2^2}\right]$. The BS employs $N=256$ antennas and $N_{\mathrm{RF}}=4$ RF chains with $G=4$ sub-arrays. The carrier frequency is 100 GHz and the number of propagation paths is $L=2$. The pilot length is set to $T_p=32$. The spatial angle region is $\theta \in (-\sqrt{3}/2,\sqrt{3}/2)$ and the user distance range is $r \in [5\,\mathrm{m},100\,\mathrm{m}]$. 

To demonstrate the effectiveness of the proposed ASSBL algorithm, we compare it with several representative baselines. Specifically, T-OMP~\cite{10634218}, P-SBL~\cite{10153711}, and P-SOMP~\cite{Cui2022_Channel} employ a polar-domain dictionary with $Q=2201$ atoms. In addition, DFT-SBL adopts a DFT dictionary with $U=256$ angular grid points, while TS-VDCE~\cite{10509715} uses a wavenumber-domain dictionary with $Q_1=512$ atoms. For the proposed adaptive dictionary, the number of angular grid points is also set to $U=256$ with an angular resolution of $\frac{2}{N}$.

\begin{figure}[!t]
	\centering
	\includegraphics[width=3.5in]{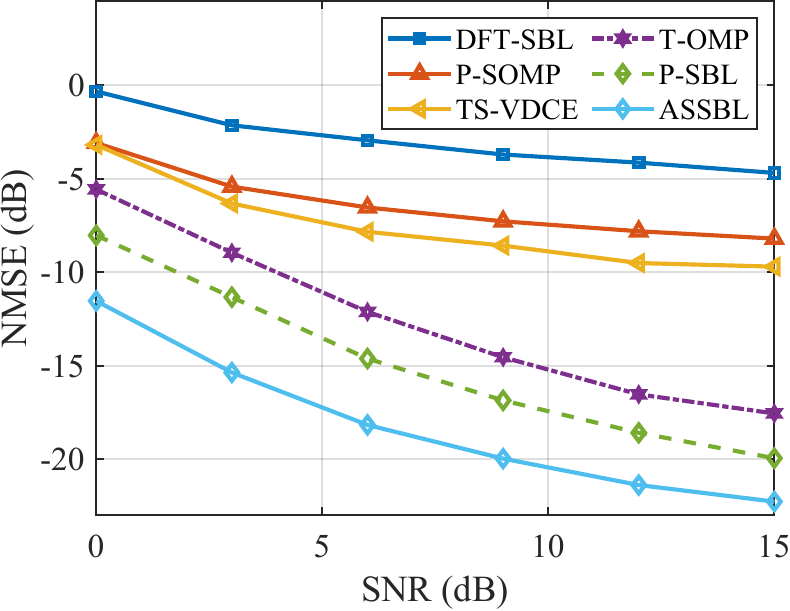}
	\caption{The NMSE performance comparison against the SNR.}
	\label{fig_2}
\end{figure}

\begin{figure}[!t]
	\centering
	\includegraphics[width=3.5in]{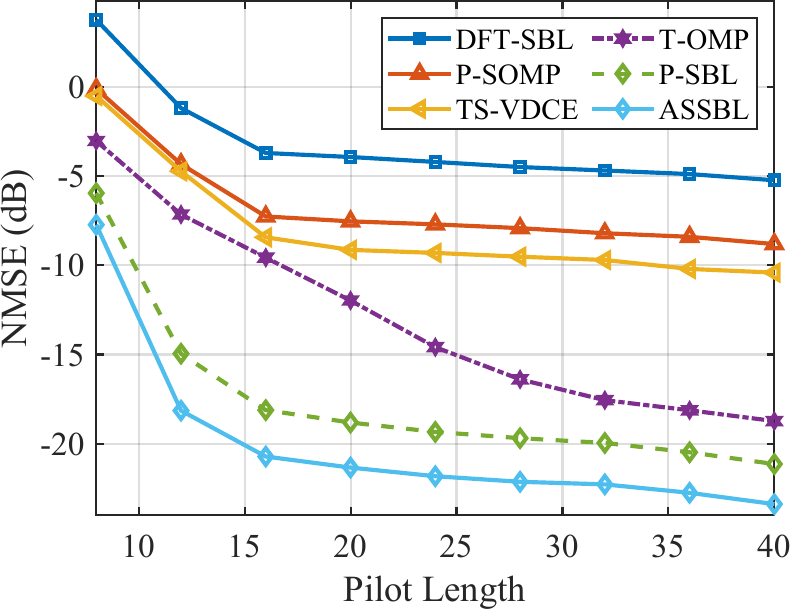}
	\caption{The NMSE performance against the length of pilot.}
	\label{fig_3}
\end{figure}
Fig. \ref{fig_2} illustrates the NMSE performance of different channel estimation algorithms versus SNR. Polar-domain-based methods generally show better performance than those relying on far-field DFT or wavenumber-domain dictionaries, since the polar-domain model captures both angular and distance-dependent features of near-field propagation. Moreover, the proposed ASSBL algorithm yields the lowest NMSE among the compared schemes under the considered settings. This performance gain stems from the hierarchical prior, which jointly exploits polar-domain sparsity and structured dependency, together with the adaptive refinement of distance parameters that mitigates grid mismatch. In contrast, the DFT-SBL method lacks explicit range information and therefore becomes less effective for near-field channel estimation. OMP-type approaches, such as P-SOMP and T-OMP, rely on greedy atom selection and are sensitive to the sparsity level assumption. The conventional P-SBL method improves robustness but lacks adaptive dictionary refinement and structured modeling of spatial non-stationarity, resulting in relatively higher NMSE compared with the proposed method. 

Fig. \ref{fig_3} depicts the NMSE performance of different channel estimation algorithms versus the pilot length $T_p$ at an SNR of $15$~dB, where $T_p$ ranges from $8$ to $40$. The NMSE of all schemes decreases as $T_p$ increases, since more pilot observations provide additional information for channel recovery. Overall, the results suggest that the proposed ASSBL framework enables reliable near-field channel estimation with low dictionary overhead, demonstrating the effectiveness of adaptive dictionary updating and hierarchical Bayesian modeling.

\section{Conclusion}
This paper investigated near-field non-stationary channel estimation for XL-MIMO systems. An adaptive structured sparse Bayesian learning framework with dictionary updating is proposed to address the dictionary overhead of conventional polar-domain methods. By iteratively refining the distance parameters, the proposed approach improves near-field channel representation without increasing the dictionary size. Moreover, a hierarchical prior is developed to jointly promote polar-domain sparsity and structured dependency, thereby facilitating efficient Bayesian inference. Simulation results demonstrate that the proposed ASSBL method achieves superior estimation performance over existing polar-domain approaches with low dictionary overhead.


\newpage

\vfill

\end{document}